\documentclass[runningheads,a4paper]{llncs}

\usepackage{amssymb}
\usepackage{graphicx}
\usepackage{epstopdf}
\usepackage{graphics}

\usepackage{geometry}
\usepackage{epsfig}
\usepackage{subfigure}
\usepackage{amsmath}
\usepackage{algorithmic}
\usepackage{graphics}
\usepackage{amssymb}
\usepackage{alltt}
\usepackage{float}
\usepackage{array}
\usepackage{xspace}
\usepackage{footmisc}
\usepackage{comment}
\usepackage{algorithm}
\usepackage{algorithmic}

\usepackage{url}
\usepackage[justification=centering]{caption}

\usepackage{url}
\urldef{\mailsa}\path|{alfred.hofmann, ursula.barth, ingrid.haas, frank.holzwarth,|
\urldef{\mailsb}\path|anna.kramer, leonie.kunz, christine.reiss, nicole.sator,|
\urldef{\mailsc}\path|erika.siebert-cole, peter.strasser, lncs}@springer.com|
\urldef{\mailsd}\path|{cyzhang,  leizhu, jlong, linshq, zyang22, 174601025}@csu.edu.cn|
\newcommand{\keywords}[1]{\par\addvspace\baselineskip
\noindent\keywordname\enspace\ignorespaces#1}

\begin{document}

\mainmatter  

\title{A hybrid index model for efficient spatio-temporal search in HBase}

\titlerunning{A hybrid index model for efficient spatio-temporal search in HBase}

%
%
\author{}
\authorrunning{A hybrid index model for efficient spatio-temporal search in HBase}

\author{Chengyuan Zhang$^\dagger$, Lei Zhu$^\dagger$, Jun Long$^\dagger$, Shuangqiao Lin$^\dagger$, Zhan Yang$^\dagger$, Wenti Huang$^\dagger$}

\institute{$^\dagger$ School of Information Science, Central South University, PR China\\
\mailsd\\
}
%



%
%

\toctitle{Lecture Notes in Computer Science}
\tocauthor{Authors' Instructions}
\maketitle

\begin{abstract}
With advances in geo-positioning technologies and geo-location services, there are a rapidly growing massive amount of spatio-temporal data collected in many applications such as location-aware devices and wireless communication, in which an object is described by its spatial location and its timestamp. Consequently, the study of spatio-temporal search which explores both geo-location information and temporal information of the data has attracted significant concern from research organizations and commercial communities. This work study the problem of spatio-temporal \emph{k}-nearest neighbors search (ST$k$NNS), which is fundamental in the spatial temporal queries. Based on HBase, a novel index structure is proposed, called \textbf{H}ybrid \textbf{S}patio-\textbf{T}emporal HBase \textbf{I}ndex (\textbf{HSTI} for short), which is carefully designed and takes both spatial and temporal information into consideration to effectively reduce the search space. Based on HSTI, an efficient algorithm is developed to deal with spatio-temporal \emph{k}-nearest neighbors search. Comprehensive experiments on real and synthetic data clearly show that HSTI is three to five times faster than the state-of-the-art technique.
\end{abstract}

\section{Introduction}
\label{intro}

Massive amount of data, which include both geo-location and temporal information, are being generated at an unparalleled scale on the Web. For example, more than 3.2 Billion comments have been posted to Facebook every day [1], while more than 400 million daily tweets containing texts and images~\cite{DBLP:journals/tip/WangLWZ17,DBLP:journals/mta/WuHZSW15,DBLP:conf/ijcai/WangZWLFP16,YangMM15,LinMM13,DBLP:conf/cikm/WangLZ13} have been generated by 140 million twitter active users[4]. Combined with the advances in location-aware devices~\cite{DBLP:journals/tkde/ZhangZZL16,DBLP:conf/icde/ZhangZZL13,DBLP:journals/pr/WuWLG18,DBLP:journals/cviu/WuWGHL18} (GPS-enabled devices, RFIDs, etc) and wireless communication~\cite{DBLP:journals/sj/LongDOL17,DBLP:conf/sigir/WangLWZZ15,DBLP:journals/tip/WangLWZ17}, spatio-temporal data storage and processing have entered a new age. There is an increasing demand to manage spatio-temporal data in many applications, such as wireless sensor networks (WSNs)~\cite{DBLP:journals/sensors/LiuLL16,DBLP:journals/pr/WuWGL18}, spatio-temporal multimedia retrieval ~\cite{DBLP:journals/tnn/WangZWLZ17,DBLP:journals/tip/WangLWZZH15,NNLS2018,LINWIVC,LINWINF}. Consequently, the study of spatio-temporal search which explores both geo-location information and temporal information of the data has attracted significant concern from research organizations and commercial communities.

This work investigates the problem of spatio-temporal \emph{k}-nearest neighbors search (ST$k$NNS), which is applied in a variety of applications, such as spatio-temporal database management systems (STDBM), location based web service, spatio-temporal information based recommending system and industrial detection system. For example, there is a pipe leakage occurs in residential area, to confirm the leakage location and leakage reason as soon as possible, the water inspector has to query the spatio-temporal data for a given region and given time interval. However, there are lots of data satisfy the given spatio-temporal constraint sometimes. Thus, the water inspector might wish to limit the query answer to $k$ nearest spatio-temporal data.

\textbf{Challenges.} There are three key challenges in spatio-temporal \emph{k}-NN query. Firstly, vast amount of data, typically in the order of TB scale or even PB scale, are uploaded to the service. Thus, the cloud storage systems, which can exploit a distributed hash table (DHT) approach to index data, should be adopted. Secondly, current cloud storage systems, such as HBase, provide a key-value store system, but they cannot maturely extend to support multi-attribute queries, which greatly restrict their application. Hence, it is important to design a transformation mechanism to convert multi-dimension value to one-dimension value. Thirdly, novel techniques need to be created to design spatio-temporal indexing scheme that supports spatial pruning and temporal pruning synchronously.

To the best of our knowledge, ~\cite{CHENXiao-yin} is the only existing work that systematically study the problem of spatio-temporal search based on HBase. STEHIX is proposed to match each spatio-temporal to relevant spatial region and time interval. Although spatial and temporal information are taken into consideration during index construction, they are subjected to two fundamental shortcomings. Firstly, separate spatial index and temporal index are built during the index construction regardless of connections between their space region or time interval. One of our key observations is that the combination of space region and time interval can significantly reduce the candidates satisfied query constraint, as a mass of unrelated data have been excluded. Hence, to achieve better performance, an indexing mechanism should integrate both spatial and temporal information. Secondly, the spatial and temporal information are fully decoupled in STEHIX which will cost lots of unnecessary I/O access, as most of false positive results are caused by small overlapping cells.

Based on the above observation, a novel index technique is proposed, namely \textbf{H}ybrid \textbf{S}patio-\textbf{T}emporal HBase \textbf{I}ndex (\textbf{HSTI} for short), to effectively organize spatio-temporal data. In brief, HSTI is a two-layered structure which follows the retrieval mechanism of HBase. In the first layer, the whole space is partitioned into equal-size cells, and a space filling curve technique, Z-order, is employed to map these two-dimensional spaces to one-dimensional sequence number. The Z-ordering values of these spaces are used as the prefix of the row key and maintained in the META table. In the second layer, to effectively partition the spatio-temporal data, a three-dimensional tree structure is designed, named Z-Octree. To further improvement the performance of Z-Octree, a minimum bounding rectangle optimization strategy is also designed to check non-fully overlapping cell, the non-fully overlapping cell need to further access, if and only if it passes the check. Comprehensive experiments demonstrate that our HSTI achieves significant improvement while comparing with previous work.

\textbf{Contributions.} The principle contributions of this work are summarized as follows.
\begin{itemize}
\item A novel Hybrid Spatio-Temporal Hbase Index is devised to deal with the problem of spatio-temporal query. As far as we know, this work is the first spatio-temporal indexing mechanism which integrates the spatial and temporal information during index construction.
\item Based on HSTI, an efficient spatio-temporal $k$-nearest neighbors query algorithm is developed.
\item Comprehensive experiments on real and synthetic datasets demonstrate that our new index achieve substantial improvements over the state-of-the-art technique.
\end{itemize}

\textbf{Roadmap.} The rest of the paper is organized as follows. Section \ref{related work} introduces related work. Section \ref{model and structure} describes the data model, and the index structure. Section \ref{knn algorithm} presents algorithms and optimization strategies for search and refinement. Extensive experiments are depicted in Section \ref{exp}. Finally, Section \ref{con} concludes the paper.

\section{Related work}
\label{related work}

With the emergence of the era of big data~\cite{DBLP:journals/pvldb/LabrinidisJ12,DBLP:journals/corr/abs-1708-02288,TC2018}, relational DBMSs are incompetent with the increasing volume of data because of low insertion rate and insufficient scalability. Therefore, to manage and process multidimensional spatial data efficiently, tree-structured spatial indices, such as R-Tree~\cite{DBLP:conf/sigmod/Guttman84}, R*-Tree~\cite{DBLP:conf/sigmod/BeckmannKSS90}, Quad-Tree~\cite{DBLP:journals/acta/FinkelB74,DBLP:journals/pr/WuWGL18}, Kd-Tree~\cite{DBLP:journals/corr/Brown14b}, are widely used in traditional DBMSs.

Recently, spatial data with temporal attribute becomes one of the largest volumes of data collected by web services. Traditional relational DBMSs can no longer handle the quantity, and thus some researchers study on many NoSQL database implementations for scaling datastores horizontally. Fox et. al~\cite{DBLP:conf/bigdataconf/FoxEHL13} presented a spatio-temporal index built on top of Accumulo to store and search spatio-temporal data sets efficiently.

To processing large volumes of data, SpatialHadoop~\cite{DBLP:journals/pvldb/EldawyM13,DBLP:conf/mm/WangLWZZ14} and Hadoop-GIS~\cite{DBLP:journals/pvldb/AjiWVLL0S13,DBLP:conf/pakdd/WangLZW14}, which are based on MapReduce, are widely used. These systems can efficiently support high-performance spatial queries. However, they cannot directly be used in real-time system, as they do not take the temporal constraints into consideration.

Spatial indices have also been extended to NoSQL-based solutions. To partition the space on top of HBase, S. Nishimura et. al~\cite{DBLP:conf/mdm/NishimuraDAA11} built a multi-dimensional index layer, called MD-HBase, by using multidimensional index structures (K-d Tree and Quad-Tree). Linearization techniques (Z-ordering~\cite{MORTON}) are used to convert multidimensional data to one dimension. However, MD-HBase does not index inner storage structure of slave nodes, and only provides an index layer in the META table. Hence, full scan operations are executed in each slave node, which reduces its efficiency.

Hsu et al.~\cite{DBLP:conf/mdm/HsuPWPL12} presented a novel key formulation scheme for spatial index in HBase, called KR+-tree. R+-tree is used to divide the data into disjoint rectangles, while gird is used for further division. Then, Hilbert-curve is exploited to encode the grid cells. During processing, KR+-tree first searches the rectangle cells, which satisfied the query constraint, in the KeyTable. Then, it finds the corresponding data according to the rectangles. However, the scan operations still need to be executed in slave nodes, because the lookup mechanism of HBase is not considered.

Zhang et al.~\cite{DBLP:conf/trustcom/ZhangZCCC14} proposed scalable spatial data storage based on HBase, called HBaseSpatial. Compared with MongoDB and MySQL, HBase has better performance while searching complex spatial data, especially searching MultiLingString and LingString data types. But this storage model can only support spatial queries in HBase.

To the best of our knowledge,~\cite{CHENXiao-yin} is the only work, which takes both spatial information and temporal information into consideration. In~\cite{CHENXiao-yin}, Chen et al. proposed a spatio-temporal index scheme, called STEHIX (Spatio-TEmporal HBase IndeX) based on HBase. However, the spatial information and temporal information are not considered as an entirety.

\section{Model and structure}
\label{model and structure}

This section first presents a description of problem, then gives an overview of storage model based on HBase, last the proposed index structure HSTI, which based on the two-level lookup mechanism of HBase, is introduced. Table ~\ref{tab:notation} summarizes the mathematical notations used throughout this paper to facilitate the discussion of our study.

\begin{table}
	\centering
    \small
	\begin{tabular}{|p{0.12\columnwidth}| p{0.56\columnwidth} |}
		\hline
		\textbf{Notation} & \textbf{Definition} \\ \hline\hline
		~$O$                & a given data set of spatio-temporal data                                \\ \hline
        ~$o_i$              & a spatio-temporal data object                                \\ \hline
	  	~$o_{id}$           & the identification of an object                                 \\ \hline
        ~$x_i$              & the longitude of a spatio-temporal data  \\ \hline
		~$y_i$              & the latitude of a spatio-temporal data                              \\ \hline
		~$t_i$              & the timestamp of a spatio-temporal data                                \\ \hline	
    	~$p$                & the location point of an object $o_i$ \\\hline
        ~$z_n$              & the value of Morton order in Z-order curve  \\\hline
        ~$L$                & the deepest level of Z-Octree \\\hline
        ~$\xi$              & the division threshold value for a node in Z-Octree \\\hline
        ~$v$                & the Morton order of a leaf node in Z-Octree                               \\ \hline	
        ~$q$                & a spatio-temporal $k$-NN search                              \\ \hline
        ~$k$                & the nearest neighbors’ number of a $k$-NN search                              \\ \hline
        ~$\delta_e$         & the Euclidean distance between two points in spatial space                                \\\hline
	\end{tabular}
    \caption{The summary of notations} \label{tab:notation}	
\end{table}

\subsection{Problem description}

\begin{definition}[\textbf{Spatio-temporal object}] \label{def:skt sktq}
A spatio-temporal object can be represented as the form $o_i$ $(o_id, x_i, y_i, t_i)$, where $o_{id}$ is the identification of the object that has a spatial location $p$, including longitude and latitude $(x_i, y_i)$ along x and y dimensions at the timestamp $t_i$.
\end{definition}

\begin{definition}[\textbf{Spatio-temporal k-nearest neighbors search}] \label{def:skt sktq}
Given a set $O$ of spatio-temporal objects, a spatio-temporal k-nearest neighbors query is denoted as $q (x_q, y_q, [t_{start}, t_{end}], k)$, where $(x_q, y_q)$ is the query spatial location and $([t_{start}, t_{end}])$ is the query temporal interval. This work aims to find a set $O(q) \subseteq O \cap |O(q)| = k$, and for each location point of $o \in O(q)$, $\delta_e(p, (x_q, y_q))$≤$\delta_e(p', (x_q, y_q))$,$ \forall p \in O(q)$, $p' \in O\O(q)$, and $p.t$, $p'.t \in [t_{start}, t_{end}]$, where the $\delta_e$ is the Euclidean distance.
\end{definition}

\subsection{HBase storage model overview}
As the common used NoSQL database, Apache HBase~\cite{DBLP:journals/tkde/ZhangZZL16} utilizes the distributed processing of the Hadoop file system (HDFS) to achieve scalability, in which the data are organized in the form of key-value pairs. A HBase cluster, which allow large-scare of data distributed storage across multiple physical servers, is usually comprised by one or more master nodes (called Masters) and several slave nodes (called RegionServers). The implementation of HBase cluster depends on the ZooKeeper.

\begin{figure}[thb]
\newskip\subfigtoppskip \subfigtopskip = -0.1cm
\centering
\includegraphics[width=.50\linewidth]{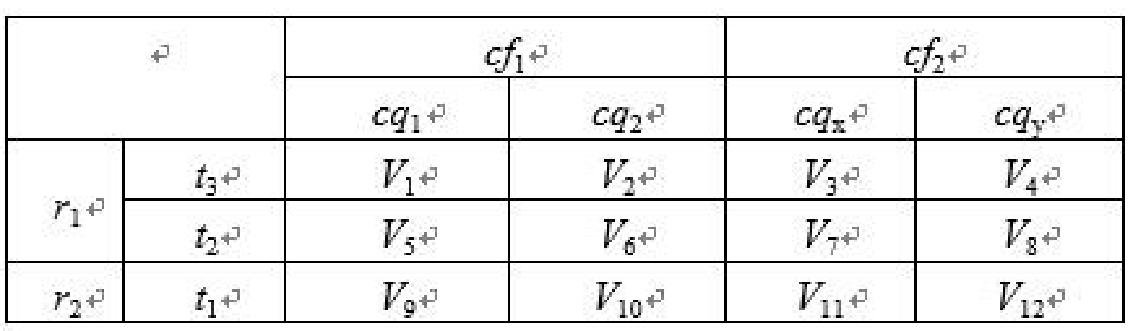}
\vspace{-1mm}
\caption{\small Logical view for HBase table }
\label{fig:table2}
\end{figure}

Unlike the data structure of traditional RDBMSs, the basic storage unit in HBase is a cell, which is defined as $<RowKey$, $Column Family$:$Column Name$, $TimeStamp>$. Fig.~\ref{fig:table2} illustrates the logical view of a table in HBase. The value $V_1$ can be retrieved by the conditions $<r_1,cf_1:cq_1,t_3>$, where $r_1$ means the row key, $cf_1$ means the column family name, $cq_1$ means column name and $t_3$ means the timestamp of this value. Within a table, rows are sorted in lexicographical order according to their unique row keys. The timestamp marks the updating of all data in the database, and each update version corresponds to a timestamp.

\begin{figure}[thb]
\newskip\subfigtoppskip \subfigtopskip = -0.1cm
\centering
\includegraphics[width=.60\linewidth]{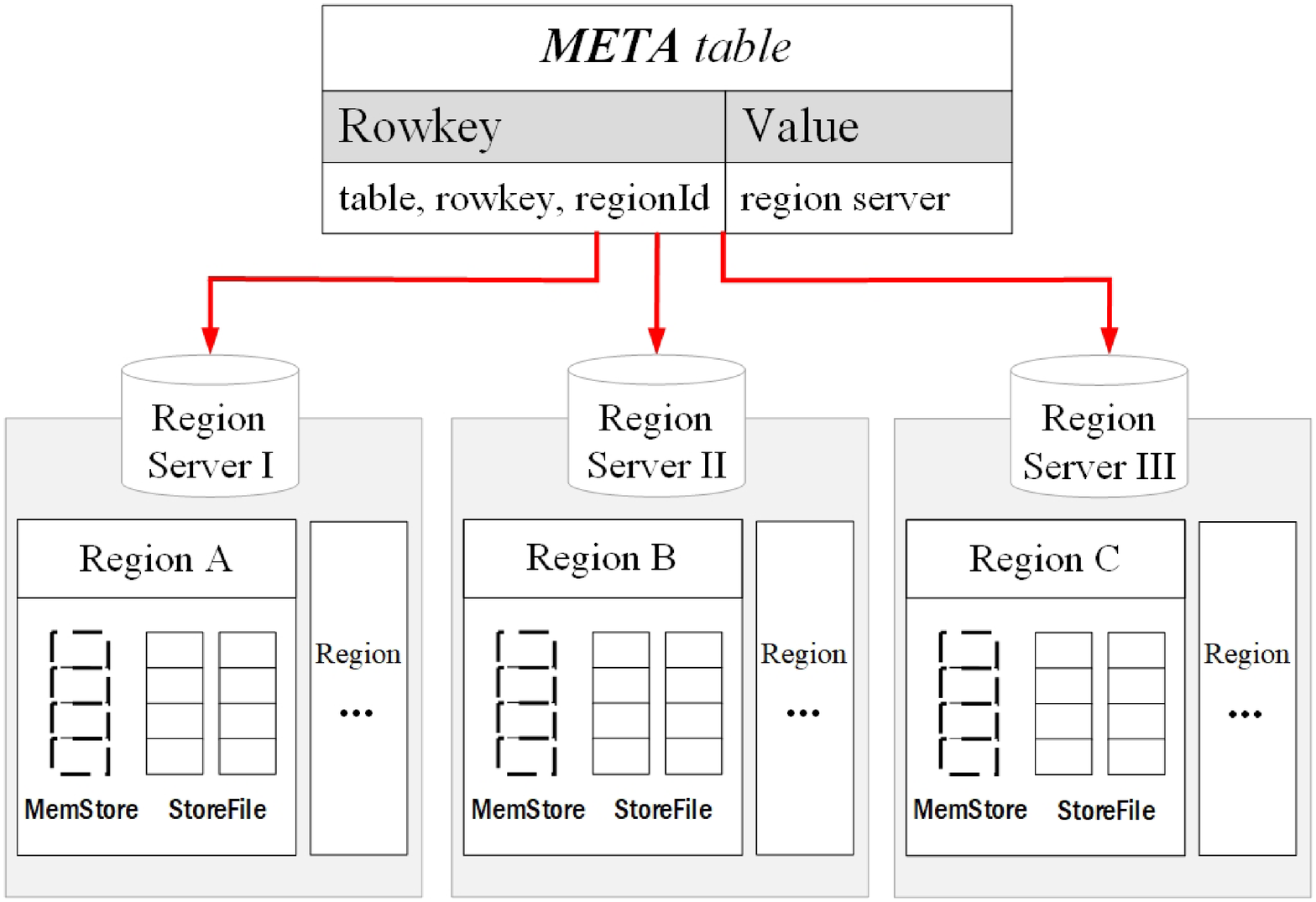}
\vspace{-1mm}
\caption{\small The two-level lookup structure in HBase }
\label{fig:Fig1}
\end{figure}

As shown in Fig.~\ref{fig:Fig1}, at the physical view for HBase, user tables are divided into several regions along their rows horizontally. Each region is maintained by exactly one RegionServer, which is the smallest unit for distributed storage and load balancing in HBase cluster. For all user tables, the location information of RegionServers was stored in META table in HBase. The data in the META table are organized as key-value pairs, where the key is $<TableName$, $RegionStartKey$, $RegionId>$ and the value is $<RegionSever>$. A key-value type of search dependences on a two-level internal lookup mechanism of HBase to locate the value. Fig.~\ref{fig:Fig1} illustrates the two-level internal lookup mechanism in HBase. The work flow is described as following, for a given row key, the exact location of the corresponding RegionServer is obtained by META table, then it will take a full scan operation in the RegionServer to find the value. For simplicity, some unrelated retrieval process, such as ROOT table search and ZooKeeper coordination, are omitted.

With the internal mechanism of HBase is further studied, we can see that there is only a simple index layer in the META table. Obviously, the full scan operation in each corresponding RegionServer is inefficient, especially when the selectivity of the spatio-temporal queries is high.

\subsection{Index structure}
\begin{figure}[thb]
\newskip\subfigtoppskip \subfigtopskip = -0.1cm
\centering
\includegraphics[width=.70\linewidth]{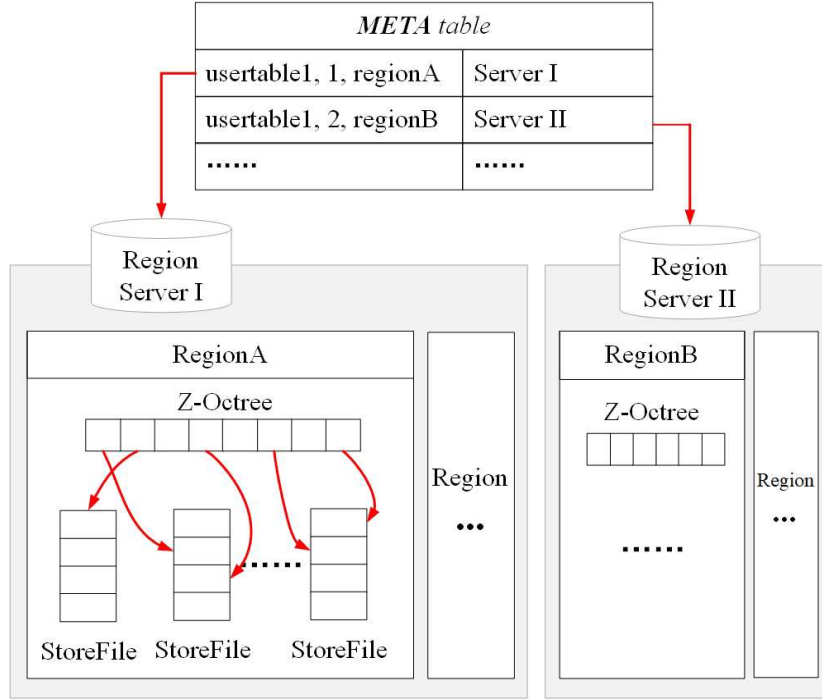}
\vspace{-1mm}
\caption{\small The overall structure of HSTI }
\label{fig:Fig2}
\end{figure}

This section introduces the novel two-layer index structure, namely HSTI, which is based on the two-level internal lookup mechanism in HBase. The first layer index in HSTI can achieve high efficiency of servers routing in HBase cluster. With the help of the second layer index in HSTI, the distributed retrievals in involved RegionServers can be completed immediately. Fig.~\ref{fig:Fig2} shows the overall structure of HSTI. The details of index design are described as below.

\subsubsection{The first layer: \emph{META} table design.}
This subsection induces the first level of HSTI. Firstly, the whole spatial space is divided into non-overlapping cells, in which spatio-temporal data points are distributed based on their spatial locations. The linearization technique Z-order curve~\cite{MORTON} is used to map data points from multi-dimension to one-dimension. As will be shown in Section 4, the reason why choosing Z-order curve is that the proposed index should have the special property: given any two-dimensional rectangle, the bottom leftmost point and the top rightmost point should be mapped as the minimum value and the maximum value respectively. All the points within the specified rectangle are distributed in this region.

Fig illustrates Z-order curves. $Z_n$($A$) is the minimum value and $Z_n$($C$) is the maximum value, with regarded to all the data within the region.

Given a data point with spatio-temporal information, the spatial Z-order curve value of the point can be calculated by its longitude and latitude $(x_i, y_i)$. According to the hash principle of HBase, the prefix of row key should be different so that it can improve the insertion rate over different RegionServers and increase the chance of load balancing. Meanwhile, the data points that are spatially close should share the same prefix in row key, thus these points can be stored adjacently by their spatial proximity. Therefore, as the first layer to index spatio-temporal data, Z-order curve value Zn is used as row key in the \emph{META} table, and then each data point can be placed into the corresponding RegionServer.


\subsubsection{The secound layer: Z-Octree structure}
In each RegionServer, the spatio-temporal data points are stored discretely. When a non-primary key search is executed in HBase, such as spatio-temporal \emph{k}-nearest neighbors search, full scan operations would be taken in regions to find \emph{k} points that satisfy query constraint. Obviously, scan operations without any optimization measure in RegionServers will degrade overall query performance in HBase. It is necessary to maintain an spatio-temporal index scheme in each RegionServer, which can avoid unnecessary scan in spatio-temporal \emph{k}-nearest neighbors search.

For searching local data efficiently, a novel indexing structure, namely Z-order Octree (Z-Octree), is proposed as the second layer in HSTI. Z-Octree is an in-memory list-like structure, kept in each RegionServer. Each entry in Z-Octree records a list of addresses pointing to spatio-temporal data in StoreFiles.


In Z-Octree, the spatio-temporal space is considered as a three-dimensional space, in which all the data points are mapped by their spatial and temporal information. In the three-dimensional data processing, octree structure is an efficient storage method so that octree is adopted here to store the code values. In octree, each non-leaf nodes have 8 sub-nodes as show in Fig.5(a). In~\cite{DBLP:journals/corr/Brown14b}, a three-dimensional space can be recursively partitioned into uniform $8^{\emph{L}-1}$ subspaces where \emph{L} is the level of the partition. Each subspace is assigned a Morton value based on its visiting order. Nevertheless, unlikely with the traditional process, a more applicable strategy is proposed to construct an octree (called Z-Octree).

To keep the Z-Octree adaptive, it is constructed based on the data dense and skewness. An inner subspace would divide into 8 subspaces only when there are sufficient points in it. Hence in our Z-Octree structure, the leaf nodes can exist on different levels which is different with traditional octree. When construct the Z-Octree, the whole spatio-temporal space is firstly processed as a root node, and then if a space contains more than $\xi$ points, it will be recursively divided into 8 subspaces. As shown in Fig(a), the whole space is divided into eight sub-nodes, and only one of them satisfies the division threshold value $\xi$. From this way, Z-Octree can easily handle the skew data and eliminate the hotspot efficiently.

The generation of Morton order value for each leaf node is described as follows. For the given deepest level \emph{L} of the octree, we assume that the whole spatio-temporal space can be divided into $8^{\emph{L}-1}$ virtual subspaces and the Morton order value of each virtual subspace is calculated~\cite{MORTON}. In Z-Octree, it does not have to implement each virtual subspace in sense that, if the number of a node does not reach the division threshold value $\xi$, this node will not be divided. The minimum of virtual subspace, which covers the not divided leaf node, is used as the Morton value \emph{v} of this specified node. Fig.5(b) shows an example of Z-Octree built from Fig.5(a) which has a deepest level 3. The Morton value of each leaf node is denoted in the corresponding circle. 



\section{\emph{K}-nearest neighbors search algorithm}
\label{knn algorithm}

A spatio-temporal \emph{k}-NN query q is defined in Subsection 3.1. Usually, searching the \emph{k}-nearest neighbors in spatio-temporal dimension is more difficult than the search of spatial \emph{k}-NN. Because the restrictions on both spatial and temporal information should be considered.

A \emph{k}-NN query algorithm based on HSTI is proposed. For a given spatial location $(x_q, y_q)$, the Z-ordering space Zn containing $(x_q, y_q)$ is computed in first layer of HSTI. Then the Z-Octree in the corresponding RegionServer is utilized to retrieve a list of data points. Meanwhile, the adjacent spaces of $Z_n$ are also computed. A priority queue $Q$ is maintained for all the points and adjacent spaces, where priority metric is the distance from location $(x_q, y_q)$ to point or Z-ordering space. The element in priority queue is constantly dequeued, either being processed to result list or being retrieved to obtain adjacent points to be enqueued, until \emph{k}-nearest neighbors are found.

\begin{algorithm}
\begin{algorithmic}[1]
\footnotesize
\caption{\bf $k$ Nearest Neighbors Search}
\label{alg:strs}

\INPUT  $x_q, y_q, t_{start}, t_{end}, k$.
\OUTPUT $S~: $list of $k$-nearest neighbors.

\STATE $S\leftarrow\emptyset$;$Z_n\leftarrow\emptyset$;$CoveringCubes\leftarrow\emptyset$;$AdjacentSpaces\leftarrow\emptyset$;$AdjacentCubes\leftarrow\emptyset$;
       $e\leftarrow\emptyset$;$AS\leftarrow\emptyset$;
\STATE $Q\leftarrow CreatePriorityQueue()$;
\STATE $Z_n\leftarrow getZorderingSpace(x_q, y_q)$;
\STATE Enqueue($Z_n$, MINDIST(($x_q$, $y_q$), $Z_n$), $Q$);

\WHILE{$Q \ne \emptyset$}
    \STATE $e \leftarrow Dequeue(Q)$;
    \IF{$e$ is typeof Zordering space}
        \STATE $AdjacentSpaces\leftarrow getAdjacentSpaces(e)$;
        \FOR{each $AdjacentSpace AS \in AdjacentSpaces$}
            \STATE Enqueue($AS$, MINDIST(($x_q$, $y_q$), $AS$), $Q$);
        \ENDFOR
        \STATE Search($e, x_q, y_q, t_{start}, t_{end}, k$);
    \ENDIF
    \IF{$e$ is typeof cube}
        \STATE $PointsSet\leftarrow getPoints(e)$;
        \FOR{each $point \in PointsSet$}
            \STATE Enqueue($point$, $\delta_e$(($x_q$, $y_q$), $point$), $Q$);
        \ENDFOR
    \ENDIF
    \IF{$e$ is typeof point}
        \STATE Add $e$ into $S$;
        \IF{$S.length = k$}
            \RETURN {$S$};
        \ENDIF
    \ENDIF
\ENDWHILE
\end{algorithmic}
\end{algorithm}

\textbf{Algorithm 1} is the \emph{k}-NN query pseudocode. The input is query $q_k =$ ($x_q$, $y_q$, $[t_{start}, t_{end}]$, $k$), and the output is the \emph{k}-nearest points of the spatial location $(x_q, y_q)$ during $[t_{start}, t_{end}]$. The basic work flow is, proximity points of location $(x_q, y_q)$ are constantly inserted into a list S until the length of the list \emph{S} reaches \emph{k}.

In line 2, a priority queue \emph{Q} is initialized to order the elements which are enqueued by the distance. The computed Z-ordering space $z_n$ is first enqueued to \emph{Q} (line 4), while the adjacent spaces of $Z_n$ are also found and enqueued where the priority metric is the distance MINDIST~\cite{DBLP:conf/sigmod/RoussopoulosKV95}. Then elements in \emph{Q} are constantly dequeued and processed in line 6. If the element is a Z-ordering space, the Procedure 3 Search (line 12) would be executed in each involved RegionServer.

Since Z-Octree is used for space partition in each RegionServer, the query of \emph{k}-nearest neighbors will be transformed into the search of Z-Octree node’s neighbors. Each leaf node in the Z-Octree corresponds to a sub-cube in the spatio-temporal space. If cubes are adjacent, then the nodes in octree are neighbors with each other. The \emph{k}-nearest neighbors of a spatial location $(x_q, y_q)$ are searched not only in the covering cube, but also in all the cubes adjacent to the cube within $[t_{start}, t_{end}]$. Fig illustrates all cubes that need to be searched in Z-Octree, where the shade cubes are covering cubes and the cubes overlapping with the dotted region are spatial adjacent cubes which are satisfying temporal predicate $[t_{start}, t_{end}]$.


In procedure 3, the covering nodes (line 4) and their spatial adjacent cubes (line 9) are computed by function \emph{getCubes}() and function \emph{getAdjacentCubes}() respectively.  We can obtain the points in the covering nodes by searching Z-Octree in line 5. Then all the points and adjacent cubes are enqueued into priority queue \emph{Q} from line 4 to 12. The minimum distance is the Euclidean distance from location $(x_q, y_q)$ to the point. If the element is an adjacent space of $Z_n$, the algorithm enqueues the all the adjacent cubes which neighboring to the location $(x_q, y_q)$ line in 15 to 18.

As shown in Algorithm 1, if the element \emph{e} dequeued from \emph{Q} is a cube, the algorithm keeps enqueue the points in the cube by the distance line in 13 to 17. Otherwise, if \emph{e} is a point, which means the element is a result, it would be added into the result list \emph{S} (line 19). The procedure described above is looped until the length of list \emph{S} reaches \emph{k}. The results of query $q_k$ are obtained in \emph{S}.

\section{Experiments}
\label{exp}

\subsection{Experiment settings and datasets}
\textbf{Experiment settings.} With the implementation of HSTI, a comprehensive experimental evaluation is conducted to verify the performance of the scheme in a real cloud environment. This work putted up an eight-node HBase cluster. In the experiments, the Zookepper was responsible for coordination and synchronization in HBase cluster, the locations of all datasets were scaled to the two-dimensional space [0, 10000][0, 10000], and the timestamp of all datasets were scaled to [0, 5000]. In addition, the top \emph{k} value changed from 10 to 500, and cluster size varied from 2 to 8. By default, top \emph{k} value and cluster size were set to 100, 4 respectively. More detailed configuration of experiments is shown in table ~\ref{tab:configurations}, where the default values of the parameters are in bold.

\begin{table}
	\centering
    \small
	\begin{tabular}{|p{0.20\columnwidth}| p{0.50\columnwidth} |}
		\hline
		\textbf{Parameter} & \textbf{Configuration} \\ \hline\hline
		~CPU                & Intel Core i5 @ 3.10GHz                                \\ \hline
        ~Memory             & 4GB                                \\ \hline
	  	~Network            & 1Mbps bandwidth                                 \\ \hline
        ~OS                 & Ubuntu (Version 14.04 LTS)  \\ \hline
		~JVM                & Java 1.8.0                              \\ \hline
		~Hadoop             & version 2.6.4                                \\ \hline	
    	~HBase              & version 1.2.4 \\\hline
        ~ZooKeeper          & version 3.4.9  \\\hline
        ~Spatial region     & \textbf{200} \\\hline
        ~Time interval      & \textbf{200} \\\hline
        ~$k$                & 10, 20, 50, \textbf{100}, 200, 500                              \\ \hline
        ~Cluster Size       & 2, \textbf{4}, 6, 8                              \\ \hline
	\end{tabular}
    \caption{The detail of configurations} \label{tab:configurations}	
\end{table}

\textbf{Datasets.} Three different datasets were used in the experiments, one of which was a synthetic uniform dataset (UN for short) generated by program, and others were real-world datasets, described as following: the first one was collected in geolife project~\cite{DBLP:journals/debu/ZhengXM10} (GL for short) by 182 users from April 2007 to August 2012, the second one was T-Drive~\cite{DBLP:conf/kdd/YuanZXS11,DBLP:conf/gis/YuanZZXXSH10} (TD for short) generated by 33 thousand taxis on Beijing road network over a period of 3 months. the important statistics of three datasets is shown in table ~\ref{tab:Dataset}.

\begin{table}
	\centering
    \small
	\begin{tabular}{|p{0.30\columnwidth}| p{0.12\columnwidth} | p{0.12\columnwidth} | p{0.12\columnwidth} |}
		\hline
		\textbf{Property} & \textbf{UN} & \textbf{GL} & \textbf{TD}\\ \hline\hline
		~Number of records (millions)     & 20 & 25.84 & 17.76 \\ \hline
        ~Size of dataset (GB)             & 0.74 & 1.54 & 0.71 \\ \hline
	\end{tabular}
    \caption{Dataset statistics} \label{tab:Dataset}	
\end{table}

For accurate analyzing and evaluating, STEHIX was chosen as baseline, which has a similar index scheme with ours propose. In the experimental, the deepest level L of Z-Octree was set to 16 and the split threshold value $\xi$ was set to 200.

\subsection{Performance evaluation}
\textbf{Evaluation on different datasets.}
We investigate the query response time, index construction time and index size of 2
algorithms against three datasets Tigers, GL, TD and UN, where other parameters are set to default values. Fig.~\ref{fig:Fig1011}(a) depicts the rate of space occupying by the index sizes. The baseline requires more space due to the two kinds of indices (called \emph{s-index} and \emph{t-index}) kept for all the entries and the storage cost of it increases faster in larger datasets. In contrast, an index record is maintained for each entry only once so that our index saves more space in memory. Fig.~\ref{fig:Fig1011}(b) shows the difference of construction time between HSTI and the baseline. Due to the simple split and code algorithms of an Z-Octree, HSTI has a shorter constructing time as compared to the baseline, which need to traverse two indices during the construction. In Fig.~\ref{fig:Fig1011}(c), HSTI demonstrates superior performance in comparison with baseline in all datasets.


\textbf{Evaluation on the effect of the number of results \emph{k}.}
Figs report the
query response time of the
algorithms as a function of \emph{k} on two datasets TD and
UN. As expected, the performance of all algorithms degrades
regarding the increase of \emph{k} (i.e., the larger search region). Compared with UN, the growth of the searching cost of HSTI is much slower for the TD dataset. And the performance of HSTI always outperforms the baseline for both high-density and low-density points. The reasons are as follow. First of all, the retrieval procedures of baseline have been divide into two parts: \emph{s-index} and \emph{t-index}. Thus, each information extraction will decompose into two processes to collect results in temporal dimension and spatial dimension, which will provoke more I/O costs. On the other hand, the periodicity of \emph{t-index} gives rise to lots of discrete timestamps are mixed together. For example, assume each cycle has 24 hours, and each cycle is divided into several segments such as 8, then all data entries will map into these 8 segments by their temporal information. For a given \emph{k}-NN query, all results returned by the baseline probably have same time intervals but not in different dates. Thus, it have to spend some time to remove false positive results, which leads to unnecessary time consumption.


\textbf{Evaluation on the effect of cluster size.}
As shown in Figs, with cluster size increased, running time decreased gradually. More clusters means less data on each sever. Obviously, the processing time will decline. Meanwhile, we also observe that the performance of \emph{k}-NN query on uniform dataset UN is better than that of real-world dataset TD. This is because the data distribution of TD might be inhomogenous, and there may be some hotpot in TD.


\section{Conclusions}
\label{con}
The problem of spatio-temporal \emph{k}-NN search is important due to the increasing amount of spatio-temporal data collected in a wide spectrum of application. The proposed hybrid spatio-temporal index scheme is based on the two-level internal lookup mechanism in HBase. Based on HSTI, an efficient algorithm is developed to support spatio-temporal \emph{k}-NN search. Our comprehensive experiments on real and synthetic data clearly show that HSTI is able to achieve a reduction of the processing time by 60-80\% compared with prior state-of-the-art methods.

\bibliographystyle{spmpsci}      

\bibliography{ref}

\begin{thebibliography}{10}
\providecommand{\url}[1]{{#1}}
\providecommand{\urlprefix}{URL }
\expandafter\ifx\csname urlstyle\endcsname\relax
  \providecommand{\doi}[1]{DOI~\discretionary{}{}{}#1}\else
  \providecommand{\doi}{DOI~\discretionary{}{}{}\begingroup
  \urlstyle{rm}\Url}\fi

\bibitem{DBLP:journals/pvldb/AjiWVLL0S13}
Aji, A., Wang, F., Vo, H., Lee, R., Liu, Q., Zhang, X., Saltz, J.H.:
  Hadoop-gis: {A} high performance spatial data warehousing system over
  mapreduce.
\newblock {PVLDB} \textbf{6}(11), 1009--1020 (2013)

\bibitem{DBLP:conf/sigmod/BeckmannKSS90}
Beckmann, N., Kriegel, H., Schneider, R., Seeger, B.: The r*-tree: An efficient
  and robust access method for points and rectangles.
\newblock In: Proceedings of the 1990 {ACM} {SIGMOD} International Conference
  on Management of Data, Atlantic City, NJ, May 23-25, 1990., pp. 322--331
  (1990)

\bibitem{DBLP:journals/corr/Brown14b}
Brown, R.A.: Building a balanced k-d tree in o(kn log n) time.
\newblock CoRR \textbf{abs/1410.5420} (2014)

\bibitem{DBLP:journals/pvldb/EldawyM13}
Eldawy, A., Mokbel, M.F.: A demonstration of spatialhadoop: An efficient
  mapreduce framework for spatial data.
\newblock {PVLDB} \textbf{6}(12), 1230--1233 (2013)

\bibitem{DBLP:journals/acta/FinkelB74}
Finkel, R.A., Bentley, J.L.: Quad trees: {A} data structure for retrieval on
  composite keys.
\newblock Acta Inf. \textbf{4}, 1--9 (1974)

\bibitem{DBLP:conf/bigdataconf/FoxEHL13}
Fox, A.D., Eichelberger, C.N., Hughes, J.N., Lyon, S.: Spatio-temporal indexing
  in non-relational distributed databases.
\newblock In: Proceedings of the 2013 {IEEE} International Conference on Big
  Data, 6-9 October 2013, Santa Clara, CA, {USA}, pp. 291--299 (2013)

\bibitem{DBLP:conf/sigmod/Guttman84}
Guttman, A.: R-trees: {A} dynamic index structure for spatial searching.
\newblock In: SIGMOD'84, Proceedings of Annual Meeting, Boston, Massachusetts,
  June 18-21, 1984, pp. 47--57 (1984)

\bibitem{DBLP:conf/mdm/HsuPWPL12}
Hsu, Y., Pan, Y., Wei, L., Peng, W., Lee, W.: Key formulation schemes for
  spatial index in cloud data managements.
\newblock In: 13th {IEEE} International Conference on Mobile Data Management,
  {MDM} 2012, Bengaluru, India, July 23-26, 2012, pp. 21--26 (2012)

\bibitem{DBLP:journals/pvldb/LabrinidisJ12}
Labrinidis, A., Jagadish, H.V.: Challenges and opportunities with big data.
\newblock {PVLDB} \textbf{5}(12), 2032--2033 (2012)

\bibitem{DBLP:journals/sensors/LiuLL16}
Liu, A., Liu, X., Long, J.: A trust-based adaptive probability marking and
  storage traceback scheme for wsns.
\newblock Sensors \textbf{16}(4), 451 (2016).
\newblock \doi{10.3390/s16040451}.
\newblock \urlprefix\url{https://doi.org/10.3390/s16040451}

\bibitem{DBLP:journals/sj/LongDOL17}
Long, J., Dong, M., Ota, K., Liu, A.: A green {TDMA} scheduling algorithm for
  prolonging lifetime in wireless sensor networks.
\newblock {IEEE} Systems Journal \textbf{11}(2), 868--877 (2017).
\newblock \doi{10.1109/JSYST.2015.2448355}.
\newblock \urlprefix\url{https://doi.org/10.1109/JSYST.2015.2448355}

\bibitem{MORTON}
M., M.G.: A computer oriented geodetic data base and a new technique in file
  sequencing.
\newblock New York: International Business Machines Company (1966)

\bibitem{DBLP:conf/mdm/NishimuraDAA11}
Nishimura, S., Das, S., Agrawal, D., {El Abbadi}, A.: Md-hbase: {A} scalable
  multi-dimensional data infrastructure for location aware services.
\newblock In: 12th {IEEE} International Conference on Mobile Data Management,
  {MDM} 2011, Lule{\aa}, Sweden, June 6-9, 2011, Volume 1, pp. 7--16 (2011)

\bibitem{DBLP:conf/sigmod/RoussopoulosKV95}
Roussopoulos, N., Kelley, S., Vincent, F.: Nearest neighbor queries.
\newblock In: Proceedings of the 1995 {ACM} {SIGMOD} International Conference
  on Management of Data, San Jose, California, May 22-25, 1995., pp. 71--79
  (1995)

\bibitem{LINWINF}
Wang, Y., Huang, X., Wu, L.: Clustering via geometric median shift over
  riemannian manifolds.
\newblock Information Sciences \textbf{220}, 292--305 (2013)

\bibitem{YangMM15}
Wang, Y., Lin, X., Wu, L., Zhang, W.: Effective multi-query expansions: Robust
  landmark retrieval.
\newblock In: ACM Multimedia, pp. 79--88 (2015)

\bibitem{DBLP:journals/tip/WangLWZ17}
Wang, Y., Lin, X., Wu, L., Zhang, W.: Effective multi-query expansions:
  Collaborative deep networks for robust landmark retrieval.
\newblock {IEEE} Trans. Image Processing \textbf{26}(3), 1393--1404 (2017).
\newblock \doi{10.1109/TIP.2017.2655449}.
\newblock \urlprefix\url{https://doi.org/10.1109/TIP.2017.2655449}

\bibitem{DBLP:conf/mm/WangLWZZ14}
Wang, Y., Lin, X., Wu, L., Zhang, W., Zhang, Q.: Exploiting correlation
  consensus: Towards subspace clustering for multi-modal data.
\newblock In: Proceedings of the {ACM} International Conference on Multimedia,
  {MM} '14, Orlando, FL, USA, November 03 - 07, 2014, pp. 981--984 (2014)

\bibitem{DBLP:conf/sigir/WangLWZZ15}
Wang, Y., Lin, X., Wu, L., Zhang, W., Zhang, Q.: {LBMCH:} learning bridging
  mapping for cross-modal hashing.
\newblock In: Proceedings of the 38th International {ACM} {SIGIR} Conference on
  Research and Development in Information Retrieval, Santiago, Chile, August
  9-13, 2015, pp. 999--1002 (2015)

\bibitem{DBLP:journals/tip/WangLWZZH15}
Wang, Y., Lin, X., Wu, L., Zhang, W., Zhang, Q., Huang, X.: Robust subspace
  clustering for multi-view data by exploiting correlation consensus.
\newblock {IEEE} Trans. Image Processing \textbf{24}(11), 3939--3949 (2015)

\bibitem{DBLP:conf/cikm/WangLZ13}
Wang, Y., Lin, X., Zhang, Q.: Towards metric fusion on multi-view data: a
  cross-view based graph random walk approach.
\newblock In: 22nd {ACM} International Conference on Information and Knowledge
  Management, CIKM'13, San Francisco, CA, USA, October 27 - November 1, 2013,
  pp. 805--810 (2013)

\bibitem{DBLP:conf/pakdd/WangLZW14}
Wang, Y., Lin, X., Zhang, Q., Wu, L.: Shifting hypergraphs by probabilistic
  voting.
\newblock In: Advances in Knowledge Discovery and Data Mining - 18th
  Pacific-Asia Conference, {PAKDD} 2014, Tainan, Taiwan, May 13-16, 2014.
  Proceedings, Part {II}, pp. 234--246 (2014)

\bibitem{DBLP:journals/corr/abs-1708-02288}
Wang, Y., Wu, L.: Beyond low-rank representations: Orthogonal clustering basis
  reconstruction with optimized graph structure for multi-view spectral
  clustering.
\newblock Neural Networks  (2018)

\bibitem{NNLS2018}
Wang, Y., Wu, L., Lin, X., Gao, J.: Multiview spectral clustering via
  structured low-rank matrix factorization.
\newblock {IEEE} Trans. Neural Networks and Learning Systems  (2018)

\bibitem{DBLP:conf/ijcai/WangZWLFP16}
Wang, Y., Zhang, W., Wu, L., Lin, X., Fang, M., Pan, S.: Iterative views
  agreement: An iterative low-rank based structured optimization method to
  multi-view spectral clustering.
\newblock In: Proceedings of the Twenty-Fifth International Joint Conference on
  Artificial Intelligence, {IJCAI} 2016, New York, NY, USA, 9-15 July 2016, pp.
  2153--2159 (2016)

\bibitem{DBLP:journals/tnn/WangZWLZ17}
Wang, Y., Zhang, W., Wu, L., Lin, X., Zhao, X.: Unsupervised metric fusion over
  multiview data by graph random walk-based cross-view diffusion.
\newblock {IEEE} Trans. Neural Netw. Learning Syst. \textbf{28}(1), 57--70
  (2017)

\bibitem{DBLP:journals/mta/WuHZSW15}
Wu, L., Huang, X., Zhang, C., Shepherd, J., Wang, Y.: An efficient framework of
  bregman divergence optimization for co-ranking images and tags in a
  heterogeneous network.
\newblock Multimedia Tools Appl. \textbf{74}(15), 5635--5660 (2015)

\bibitem{LINWIVC}
Wu, L., Wang, Y.: Robust hashing for multi-view data: Jointly learning low-rank
  kernelized similarity consensus and hash functions.
\newblock Image and Vision Computing \textbf{57}, 58--66 (2016)

\bibitem{DBLP:journals/pr/WuWGL18}
Wu, L., Wang, Y., Gao, J., Li, X.: Deep adaptive feature embedding with local
  sample distributions for person re-identification.
\newblock Pattern Recognition \textbf{73}, 275--288 (2018)

\bibitem{DBLP:journals/cviu/WuWGHL18}
Wu, L., Wang, Y., Ge, Z., Hu, Q., Li, X.: Structured deep hashing with
  convolutional neural networks for fast person re-identification.
\newblock Computer Vision and Image Understanding \textbf{167}, 63--73 (2018)

\bibitem{TC2018}
Wu, L., Wang, Y., Li, X., Gao, J.: Deep attention-based spatially recursive
  networks for fine-grained visual recognition.
\newblock {IEEE} Trans. Cybernetics  (2018)

\bibitem{DBLP:journals/pr/WuWLG18}
Wu, L., Wang, Y., Li, X., Gao, J.: What-and-where to match: Deep spatially
  multiplicative integration networks for person re-identification.
\newblock Pattern Recognition \textbf{76}, 727--738 (2018)

\bibitem{LinMM13}
Wu, L., Wang, Y., Shepherd, J.: Efficient image and tag co-ranking: a bregman
  divergence optimization method.
\newblock In: ACM Multimedia (2013)

\bibitem{CHENXiao-yin}
Xiao-yin, C., Chong, Z., Bin, G., Wei-dong, X.: Efficient historical query in
  hbase for spatio-temporal decision support.
\newblock International Journal of Computers, Communications and Control
  \textbf{11}(5), 613--630 (2016)

\bibitem{DBLP:conf/kdd/YuanZXS11}
Yuan, J., Zheng, Y., Xie, X., Sun, G.: Driving with knowledge from the physical
  world.
\newblock In: Proceedings of the 17th {ACM} {SIGKDD} International Conference
  on Knowledge Discovery and Data Mining, San Diego, CA, USA, August 21-24,
  2011, pp. 316--324 (2011)

\bibitem{DBLP:conf/gis/YuanZZXXSH10}
Yuan, J., Zheng, Y., Zhang, C., Xie, W., Xie, X., Sun, G., Huang, Y.: T-drive:
  driving directions based on taxi trajectories.
\newblock In: 18th {ACM} {SIGSPATIAL} International Symposium on Advances in
  Geographic Information Systems, {ACM-GIS} 2010, November 3-5, 2010, San Jose,
  CA, USA, Proceedings, pp. 99--108 (2010)

\bibitem{DBLP:conf/icde/ZhangZZL13}
Zhang, C., Zhang, Y., Zhang, W., Lin, X.: Inverted linear quadtree: Efficient
  top k spatial keyword search.
\newblock In: 29th {IEEE} International Conference on Data Engineering, {ICDE}
  2013, Brisbane, Australia, April 8-12, 2013, pp. 901--912 (2013)

\bibitem{DBLP:journals/tkde/ZhangZZL16}
Zhang, C., Zhang, Y., Zhang, W., Lin, X.: Inverted linear quadtree: Efficient
  top {K} spatial keyword search.
\newblock {IEEE} Trans. Knowl. Data Eng. \textbf{28}(7), 1706--1721 (2016)

\bibitem{DBLP:conf/trustcom/ZhangZCCC14}
Zhang, N., Zheng, G., Chen, H., Chen, J., Chen, X.: Hbasespatial: {A} scalable
  spatial data storage based on hbase.
\newblock In: 13th {IEEE} International Conference on Trust, Security and
  Privacy in Computing and Communications, TrustCom 2014, Beijing, China,
  September 24-26, 2014, pp. 644--651 (2014)

\bibitem{DBLP:journals/debu/ZhengXM10}
Zheng, Y., Xie, X., Ma, W.: Geolife: {A} collaborative social networking
  service among user, location and trajectory.
\newblock {IEEE} Data Eng. Bull. \textbf{33}(2), 32--39 (2010)

\end{thebibliography}

\end{document}